# Upper Critical Fields and Anisotropy of Ca$_{1-x}$La$_x$Fe$_2$As$_2$ Single Crystals

W. Zhou, F. F. Yuan, J. C. Zhuang, Y. Sun, Y. Ding, L. J. Cui, J. Bai, Z. X. Shi[a)]

*Department of Physics, Southeast University, Nanjing, 211189, China*

## Abstract

Superconductivity of Ca$_{1-x}$La$_x$Fe$_2$As$_2$ single crystals with various doping level were investigated via electromagnetic measurements for out-plane ($H//c$) and in-plane ($H//ab$) directions. Obvious double superconducting transitions, which can survive in magnetic fields up to several Tesla, were observed in the medium-doped ($x = 0.13$) sample. Two kinds of distinct $H_{c2}$ phase diagrams were established for the low superconducting phase with $T_c$ lower than 15 K and the high superconducting phase with $T_c$ of over 40 K, respectively. Both the two kinds of phase diagrams exist in the medium-doped sample. Unusual upward curvature near $T_c$ was observed in $H_{c2}$ phase diagrams and analyzed in detail. Temperature dependences of anisotropy for different doping concentrations were obtained and compared. Both superconducting phases manifest extremely large anisotropies, which may originate from the interface or intercalation superconductivity.

# 1. INTRODUCTION

The discovery of iron-based superconductors (IBSs) has triggered another worldwide research to explore the mechanism of high temperature superconductivity [1]. These new-found IBSs, mainly composed by "1111" (LaFeAs(O,F)) [1], "122" ((Ba,K)Fe$_2$As$_2$) [2], "111" (LiFeAs) [3], and "11" (Fe(Se,Te)) [4] systems, share many primary characters with cuprates, such as the layer crystal structure [1], AFM parent phase [5], and high upper critical field [6]. As an important parameter for both exploring the superconductivity mechanism and the potential application, $H_{c2}$ and anisotropy were widely investigated in cuprates as well as IBSs. Many cuprate superconductors exhibit high $H_{c2}$ and significant anisotropy [7, 8] which relates to the layered structure with quasi-two-dimensional electronic properties. Early on, large $H_{c2}$ exceeding 100 T with modest anisotropy of 4-5 was found in iron pnictides [9, 10]. More interestingly, subsequent researches reported nearly isotropic $H_{c2}$ in many iron pnictides at low temperatures [11-14], indicating more three dimensional electron structure than that of cuprates. The $H_{c2}(T)$ curves of iron pnictides also exhibit complicated and puzzling upward or downward curvatures which were ascribed to multiband effect in extensive studies [15-18].

Since the discovery of Ca$_{1-x}$RE$_x$Fe$_2$As$_2$ compounds, various experiments, like pressure effect [19], NMR [20], and ARPES study [21], have been carried out on Ca$_{1-x}$RE$_x$Fe$_2$As$_2$ single crystals for the reason that those compounds bear the highest $T_c$ over 40 K among the discovered 122 type IBSs. Until now, only few work has been performed on $H_{c2}$ and anisotropy of Ca$_{1-x}$RE$_x$Fe$_2$As$_2$ [22, 23] even though it possesses the potential prospect for application. To make clear the primary superconducting parameters in the electron-doped Ca$_{1-x}$RE$_x$Fe$_2$As$_2$ compounds, systematic study of $H_{C2}$ and anisotropy of Ca$_{1-x}$RE$_x$Fe$_2$As$_2$ is remarkably meaningful. In this paper, we report measurements of magneto-resistivity on Ca$_{1-x}$La$_x$Fe$_2$As$_2$ ($x$ = 0.08, 0.13, 0.21, 0.23) single crystals. The $H_{c2}$ values for out-plane ($H//c$) and in-plane ($H//ab$) directions were obtained. Obvious double superconducting transitions with robust superconductivity against magnetic fields were found in Ca$_{0.87}$La$_{0.13}$Fe$_2$As$_2$ and the $H_{c2}$ phase diagram of the low superconducting transition was first established. Two kinds of distinct phase diagrams and unusual upward curvature near $T_c$ were observed and discussed. Temperature dependence of anisotropy parameters for different doping concentrations were obtained and analyzed.

# 2. EXPERIMENT

Ca$_{1-x}$RE$_x$Fe$_2$As$_2$ single crystals were grown by FeAs flux method. The FeAs precursor was first synthesized by reacting stoichiometric amounts of Fe and As at 750 °C for 24 h in a vacuum quartz tube. Ca, La and FeAs with certain mole ratio were placed in an alumina crucible and sealed in vacuum quartz tubes with pressure inside around $8.5\times10^{-5}$ mbar. The samples were quickly heated to 1180 °C and held for 2 h, then slowly cooled down to 970 °C with rate of -3 ~ 6 °C/h to grow crystal. The obtained crystals were easily cleaved into shining plates along the crystalline $c$-axis direction. Details of structure characterization and composition analyses of $Ca_{1-x}La_xFe_2As_2$ single crystals will be reported elsewhere [24]. Only (00$l$) peaks were observed in X-ray diffraction (XRD) patterns, demonstrating good $c$-axis orientation of crystal. The actual doping concentration $x$ determined by averaging the energy dispersive X-ray spectroscope (EDX) measurement results of several points on each crystal was used in this report. Resistivity measurements were carried out through four-probe method and the probes were attached to the $ab$ plane of crystal by using silver epoxy. Transport measurements were performed on a physical property measurement system (PPMS, Quantum Design). All crystals used for different measurements were come from same batch of samples prepared by the method stated above.

## 3. RESULTS AND DISCUSSION

Figure 1 shows temperature dependences of resistivity for the four samples ($Ca_{1-x}La_xFe_2As_2$, $x$ = 0.08, 0.13, 0.21, 0.23) used in this report. According to the previous studies of $Ca_{1-x}La_xFe_2As_2$ [25, 26], the La concentration $x$ of the four samples locates in three typical doping levels: the low ($x$ = 0.08), medium ($x$ = 0.13), and high ($x$ = 0.21, 0.23) doping level, respectively. The $RT$ curve of the low-doped sample ($x$ = 0.08) exhibits both a SDW/structure transition [25] around 115 K and a low superconducting transition below 14 K. For the medium-doped sample ($x$ = 0.13), SDW/structure transition disappears, indicating the suppression of SDW/structure transition with increasing doping. On the other hand, the $RT$ curve shows two superconducting transitions, which may relate to the coexistence of two separate superconducting phases as suggested by Lv *et al*. [27] The high superconducting transition with $T_c$ over 40 K is higher than any other known 122 type IBSs [2, 28, 29] and close to the previous report [30]. The low transition $T_c$ is just around 10 K, which can be seen more clearly in figure 3. For the high-doped samples ($x$ = 0.21, 0.23), the $RT$ curves show only high temperature superconducting transitions.

Temperature dependences of normalized resistivity in magnetic fields for three typical doping levels ($x$ = 0.08, 0.13, 0.23) are shown in figure 2. For all studied samples, superconductivity is suppressed in fields for both directions. The suppression of superconductivity with magnetic fields applied in out-plane ($H//c$) direction is more obvious than that in in-plane direction ($H//ab$), especially for the low-doped sample. In figure 3, we show the enlarged view of temperature-dependent magneto-resistivity of $Ca_{0.87}La_{0.13}Fe_2As_2$ to see the low superconducting transition more clearly. As can be seen, within our measured temperature range, superconductivity can easily survive in magnetic fields up to 9 T for the in-plane direction ($H//ab$), but was completely suppressed in magnetic fields around 3 T for the out-plane direction ($H//c$), indicating a large anisotropy for the low superconducting transition.

It should be noted here, compared with other 122 type Fe-based single crystals, superconducting transitions for $Ca_{1-x}La_xFe_2As_2$ are relatively broad for $RT$ measurement [27], which adds the difficulty in accurate determination of $H_{c2}$. Therefore, three different criteria were adopted to determine the $H_{c2}$ values, namely the onset point of the peak in $\partial\rho_{normalized}/\partial T$ curve (onset), 90% of normal state resistivity (90%), and the peak point in $\partial\rho_{normalized}/\partial T$ curve (peak). Here $\rho_{normalized} = \rho/\rho_{45K}$ or $\rho/\rho_{20K}$ according to $T_c$ values of different samples. Next, we will use the experimental data of $Ca_{0.77}La_{0.23}Fe_2As_2$ as an example to show the process of obtaining $H_{c2}$ and do some analysis. Figure 4A shows the experimental obtained $H_{c2}^c(T)$ of $Ca_{0.77}La_{0.23}Fe_2As_2$. The $H_{c2}(T)$ curves show remarkable upward curvatures near $T_c$ irrespective of $H_{c2}$ determination criteria, consistent with the previous reports [23, 30]. Since the superconducting transitions are broad for the present system, it's natural to suspect that the upward curvature may result from the random distribution of $T_c$ because of the inhomogeneity of $Ca_{1-x}La_xFe_2As_2$. Accordingly, we simply simulate the superconducting transitions in $RT$ curves for a quasi-one-dimension current path consisting of superconducting grains with different $T_c$s'. The $T_c$ values are assumed to follow the Gaussian distribution (or normal distribution), which means $T_c$ is a random variate satisfying Gaussian distribution formula $f(x) = \frac{1}{\sqrt{2\pi}\sigma}e^{-\frac{(x-\mu)^2}{2\sigma^2}}$. Here $x$ stands for $T_c$, $\mu$ is the mean and $\sigma$ is the standard deviation. $\mu$ and $\sigma$ are determined by comparison with the peak and onset transition temperature in experiment obtained $\partial(\rho/\rho_{45K})/\partial T$ curve under zero field. And for each superconducting grains, $H_{c2}$ near $T_c$ satisfies the relation suggested in Ref. [31], $H_{c2} = 4(T_c - T)\Phi_0/\pi^2 D$, where D is a constant representing the band diffusivity and $\Phi_0$ is the flux quantum. Under each magnetic field, $T_c$ was reassigned. Above $T_c$, the resistivity of each grain follows the same temperature dependence obtained through fitting of our experimental data. The total resistivity is sum of every grain. The simulating results are shown in figure 4B (inset gives the simulating $\partial(\rho/\rho_{45K})/\partial T$ curves). No upward curvature is found. Therefore, the upward curvature cannot be aroused by the broad distribution of $T_c$.

As stated above, many IBSs exhibited strange curvatures in $H_{c2}(T)$ curves and were explained well by two-band theory. To understand the upward curvature better, we also fitted our data using the two-band theory [31]. By fitting, one can obtain the band diffusivities $D_1$, $D_2$ and the $\lambda$ matrix ($\lambda = \begin{pmatrix} \lambda_{11} & \lambda_{12} \\ \lambda_{21} & \lambda_{22} \end{pmatrix}$, $\lambda_{11}$, $\lambda_{22}$ and $\lambda_{12}$, $\lambda_{21}$ are intraband- and interband- coupling constants and the exact relations can be found in Ref. [31]). As shown by the solid lines in figure 5, fitting of our out-plane data for $Ca_{0.77}La_{0.23}Fe_2As_2$ resulted in $\lambda = \begin{pmatrix} 0.8 & 0.243 \\ 0.243 & 0.3 \end{pmatrix}$ and $\eta = 500$. $\eta$ is the band diffusivity ratio, that is $\eta = D_2/D_1$. However, using the same $\lambda$ matrix and tuning the values of D and $\eta$ can't obtain as good fitting results for the in-plane data and the fitting obtained $H_{c2}(0)$ is unbelievable high. Therefore, such upward curvature near $T_c$ can't simply be ascribed to multiband-superconductivity in the case of $Ca_{1-x}La_xFe_2As_2$.

In figure 6, $H_{c2}$ phase diagrams obtained by peak criterion (except for the low superconducting transition in the medium-doped sample with 90% criterion adopted) for different doping concentrations were shown. As can be seen in figure 6B, the medium-doped sample with double superconducting transitions exhibits two distinct $H_{c2}$ phase diagrams. For the low superconducting transition, $H_{c2}^c$ almost shows similar linear temperature dependence with that of the low-doped sample, indicating probably the same origin of superconductivity. For the high superconducting transition in $Ca_{0.87}La_{0.13}Fe_2As_2$ as well as that of high-doped samples, both $H_{c2}^c(T)$ and $H_{c2}^{ab}(T)$ exhibit upward curvature near $T_c$. In fact, upward curvature near $T_c$ is common in IBSs [10, 15, 16, 32]. However, there are few reports of upward curvatures for both in-plane and out-plane directions (seeing in figure 6). Nevertheless, the usual WHH linear temperature dependences of $H_{c2}$ were observed in fields beyond 1 T. Therefore, we fitted our experimental data (peak criterion) of all samples based on simple WHH model without considering the spin paramagnetic effect. $H_{c2}(0)$s' estimated by WHH model and corresponding superconducting parameters were shown in table 1. The dashed lines in figure 5 show the WHH fitting results of $Ca_{0.77}La_{0.23}Fe_2As_2$ as an example. WHH fitting seems reasonable for $Ca_{1-x}La_xFe_2As_2$ system in the medium-field range. In figure 7, temperature-dependent anisotropy parameters $\Gamma$ for different doping concentrations were shown. The anisotropy ratios show similar temperature dependence, increasing initially from $T_c$ and then tending to decrease, for all samples. The decreasing tendency of anisotropy may be caused by stronger spin paramagnetic effect in in-plane direction as the case in many IBSs [11-13, 33]. However, the $\Gamma(0)$ values (represented by empty circles) estimated through WHH model are clearly different for the low-doped sample with low $T_c$ and the medium- or high-doped samples with high superconducting transitions, which may relate to two different superconducting phases as mentioned above. $\Gamma(0)$ values of high superconducting transitions are smaller than 5, which is compatible with that of "1111" type compounds [9, 10]. However, to obtain accurate $H_{c2}(0)$s' and corresponding $\Gamma(0)$s', high-filed experiment needs to be carried out. Another phenomenon should be noted is that remarkably large anisotropy, higher than that of any other known IBSs, emerges in the present system. As is well known, $H_{c2}$ anisotropy is closely relate to the ratio of coherence lengths in different directions ($\Gamma = H_{C2}^{ab}/H_{C2}^{c} = \xi_{ab}/\xi_{c}$, $\xi_{ab}$ and $\xi_c$ are the coherence lengths of the $c$-axis and $ab$-plane). Such large anisotropy ratio, induced by the upward curvatures in $H_{c2}(T)$ curves, means either extremely large $\xi_{ab}$ or remarkably small $\xi_c$. Thus, the large anisotropy of the present system may reflect characters of certain superconductivity with special configuration, such as interface superconductivity or intercalation superconductivity. Actually, upward curvature in $H_{c2}(T)$ curve near $T_c$ is common in the case of filamentary superconductivity [32].

To date, superconducting transitions in $Ca_{1-x}RE_xFe_2As_2$ are still mysterious. From figure 3 one can see, beyond the low transition, the resistivity almost remains temperature independent for a broad temperature range to the high superconducting transition. Such a broad range of nearly temperature-independent resistivity suggests that two superconducting phases exist of high possibility. Considering the large anisotropy, it's natural to suppose that superconductivity may only locate in the interface or intercalation places where doping is effective, forming numbers of superconducting sheets. Theses superconducting sheets consisting by low and high superconducting phases spread along the $c$-axis direction and take only small percentage of the whole crystal. It should be pointed out that, the $RT$ curve of the low-doped sample $Ca_{0.92}La_{0.08}Fe_2As_2$ also shows two stages of resistivity decrease. Whether the double-phase hypothesis in the medium-doped sample is suitable to the low-doped sample is not clear.

## 4. CONCLUSION

In conclusion, we have investigated the temperature-dependent resistivity for $Ca_{1-x}La_xFe_2As_2$ ($x$ = 0.08, 0.13, 0.21, 0.23) single crystals in different magnetic fields. $H_{c2}$ for both the in-plane ($H//ab$) and out-plane ($H//c$) directions were obtained. Using simple numerical simulation, we negated the idea that the broad distribution of $T_c$ induced the upward curvature in $H_{c2}(T)$ curves. Two-band theory was also demonstrated invalidly to explain the upward curvature. Based on WHH model, $H_{c2}(0)$s' were roughly estimated and corresponding superconducting parameters were given. Temperature-dependent anisotropy ratios $\Gamma$ for different doping concentrations were calculated and found to possess extremely large values which may originate from interface or intercalation superconductivity.

**Acknowledgments**: This work was supported by the Natural Science Foundation of China, the Ministry of Science and Technology of China (973 project: No. 2011CBA00105), and Scientific Innovation Research Foundation of College Graduate in Jiangsu Province (CXZZ_0135).

**Figure Captions**

Figure 1  Temperature dependences of resistivity for $Ca_{1-x}La_xFe_2As_2$ compounds.

Figure 2  Temperature dependences of normalized resistivity in different magnetic fields for three typical doping levels in $Ca_{1-x}La_xFe_2As_2$ ($x$ = 0.08, 0.13, 0.23) single crystals. a), c), e): out-plane direction ($H//c$). b), d), f): in-plane direction ($H//ab$).

Figure 3  The enlarged view of temperature-dependent resistivity of the medium-doped sample ($x$ = 0.13).

Figure 4  The experimental (A) and simulating (B) $H_{c2}^c(T)$ of $Ca_{0.77}La_{0.23}Fe_2As_2$ determined by three different criteria: the onset point of the peak in $\partial(\rho/\rho_{45K})/\partial T$ curve (onset), 90% of normal state resistivity (90%), and the peak point in $\partial(\rho/\rho_{45K})/\partial T$ curve (peak). The solid lines are guides to the eye. Inset: experimental (A) and simulating (B) $\partial(\rho/\rho_{45K})/\partial T$ curves in different magnetic fields.

Figure 5  Two-band and WHH fittings of $H_{c2}(T)$ of $Ca_{0.77}La_{0.23}Fe_2As_2$ using the $H_{c2}$s' determined by peak criterion.

Figure 6  $H_{C2}$ phase diagrams for $Ca_{1-x}La_xFe_2As_2$ ($x$ = 0.08, 0.13, 0.21, 0.23). The data was obtained by the peak criterion except that the low superconducting transition in $Ca_{0.87}La_{0.13}Fe_2As_2$ adopted the 90% criterion. The solid lines are guides to the eye.

Figure 7  Temperature (normalized by $T_c$ values of different samples) dependences of anisotropy parameter $\Gamma$ for different doping concentrations.

Table 1 Superconducting parameters of $Ca_{1-x}La_xFe_2As_2$ single crystals obtained through WHH analysis of $H_{c2}(T)$.

| $Ca_{1-x}La_xFe_2As_2$ | $x = 0.08$ | $x = 0.13$(low) | $x = 0.13$(high) | $x = 0.21$ | $x = 0.23$ |
|---|---|---|---|---|---|
| $\mu_0 H_{c2}^{ab}(0)$ | 103.8 (T) | 28.82 (T) | 50.2 (T) | 77.1 (T) | 102.7 (T) |
| $\mu_0 H_{c2}^{c}(0)$ | 3.8 (T) | 1.94 (T) | 19.3 (T) | 18.9 (T) | 16.8 (T) |
| $dH_{c2}^{ab}/dT$ | -13.43 (T/K) | -4.18 (T/K) | -0.82 (T/K) | -3.42 (T/K) | -3.13 (T/K) |
| $dH_{c2}^{c}/dT$ | -0.55 (T/K) | -0.28 (T/K) | -1.49 (T/K) | -0.95 (T/K) | -0.95 (T/K) |
| $\xi_{ab}$ | 93.1 (Å) | 130.3 (Å) | 41.3 (Å) | 41.8 (Å) | 44.3 (Å) |
| $\xi_c$ | 3.4 (Å) | 8.8 (Å) | 15.9 (Å) | 10.2 (Å) | 7.2 (Å) |

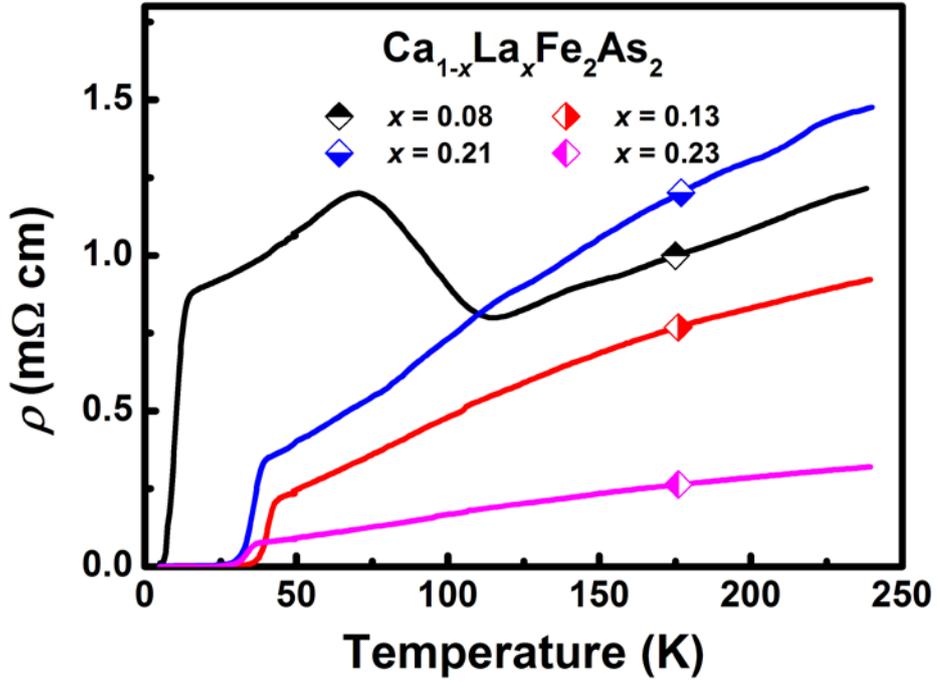

Figure 1 W. Zhou *et al*.

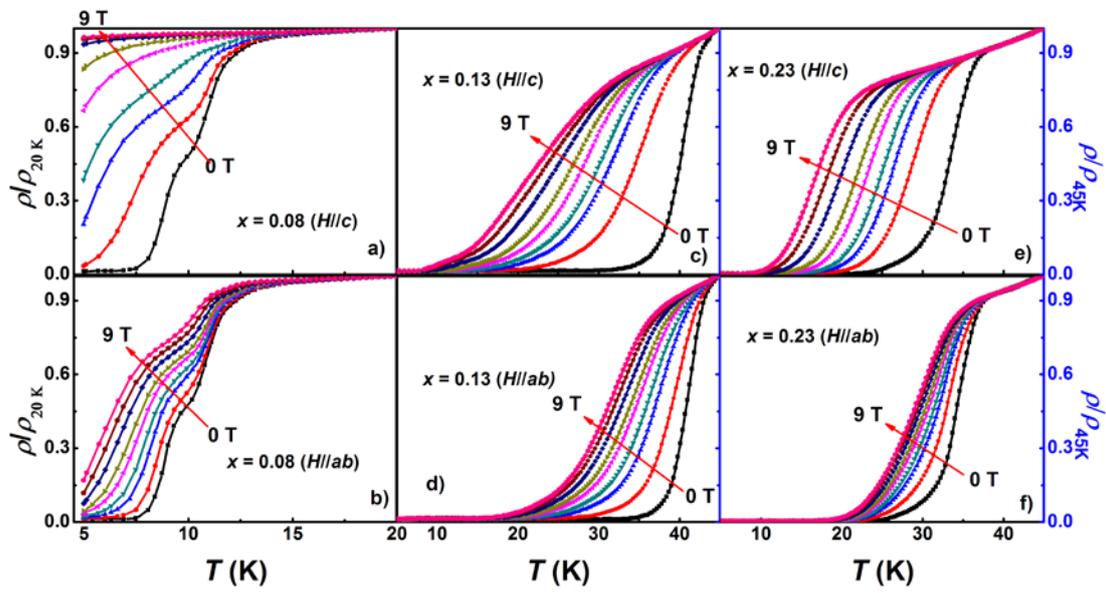

Figure 2 W. Zhou *et al*.

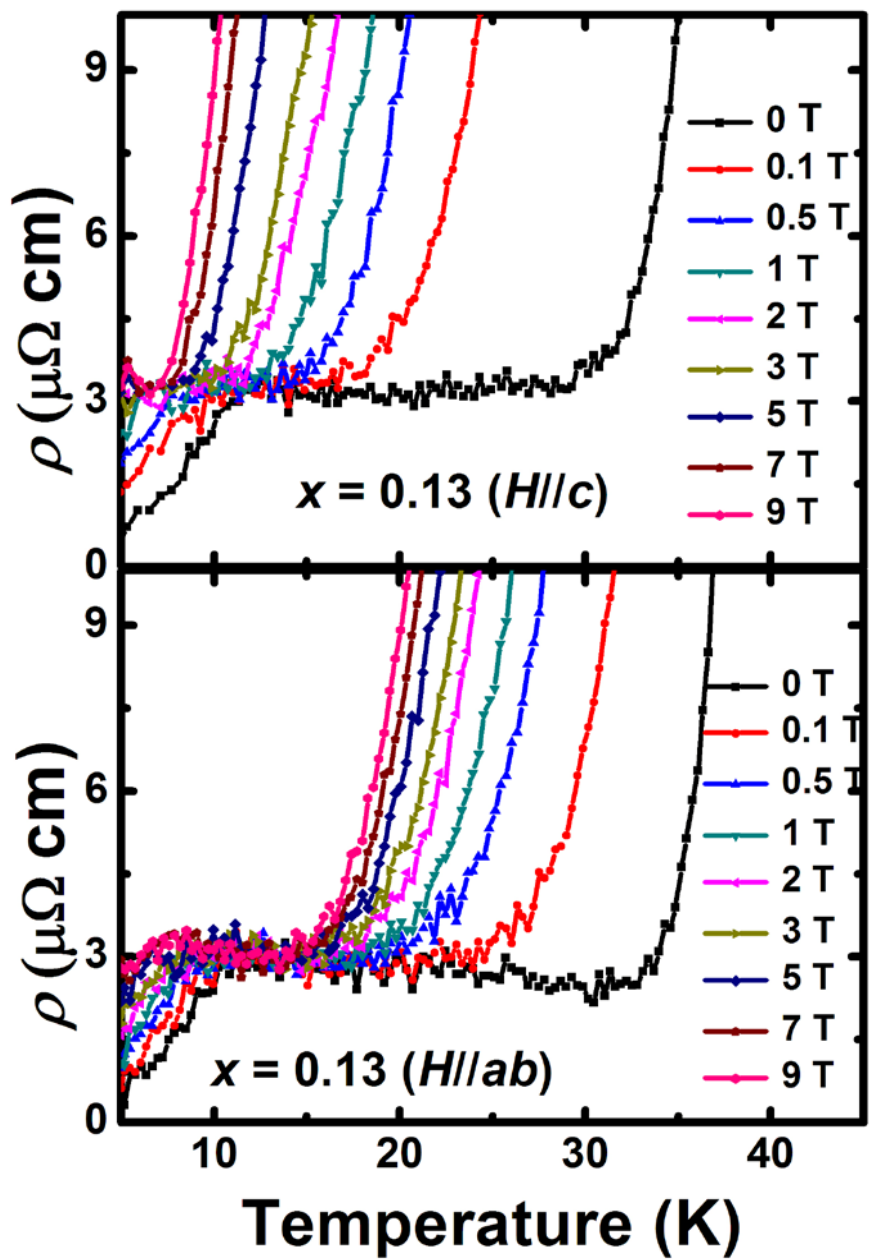

Figure 3 W. Zhou *et al*.

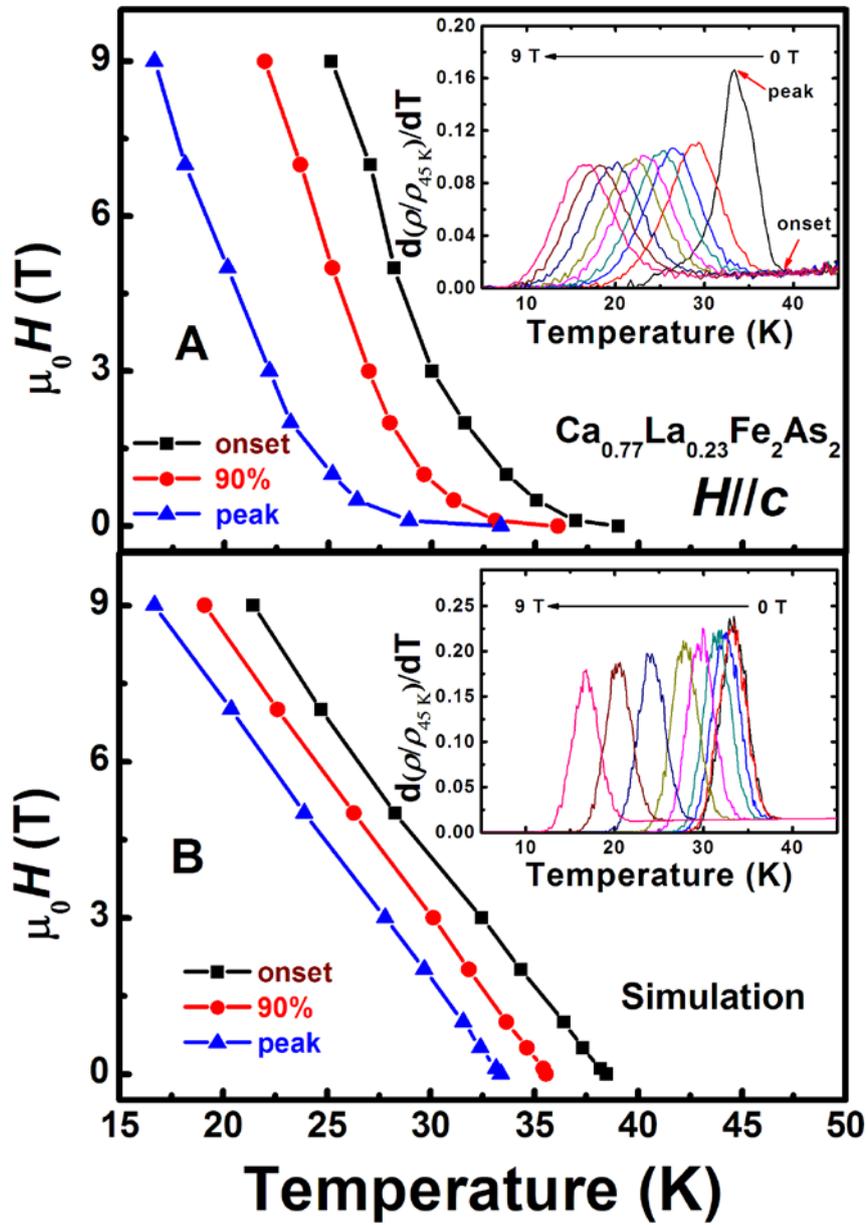

Figure 4 W. Zhou *et al*.

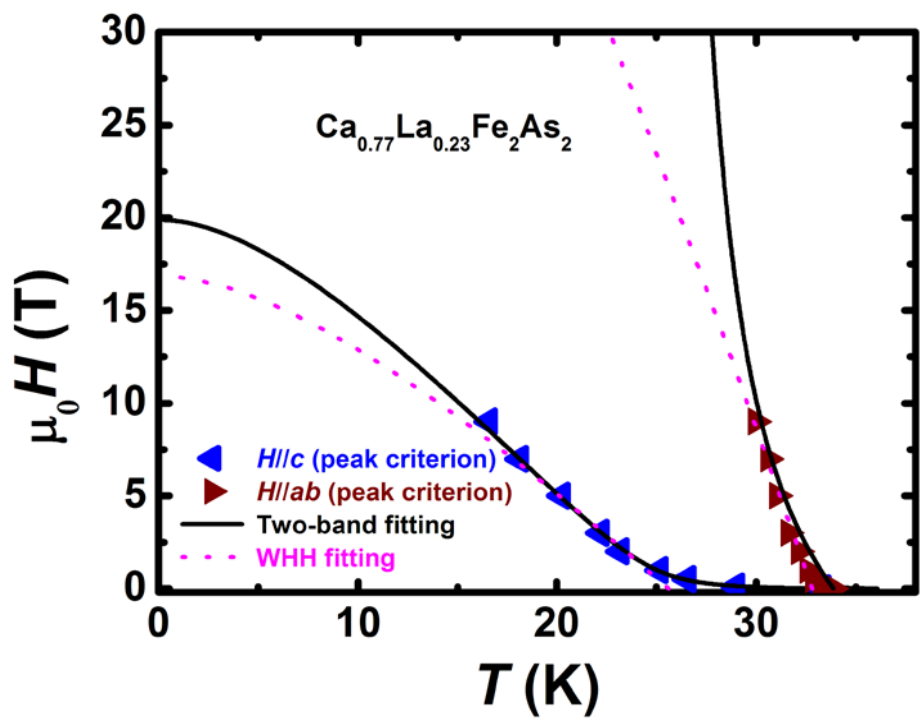

Figure 5 W. Zhou *et al*.

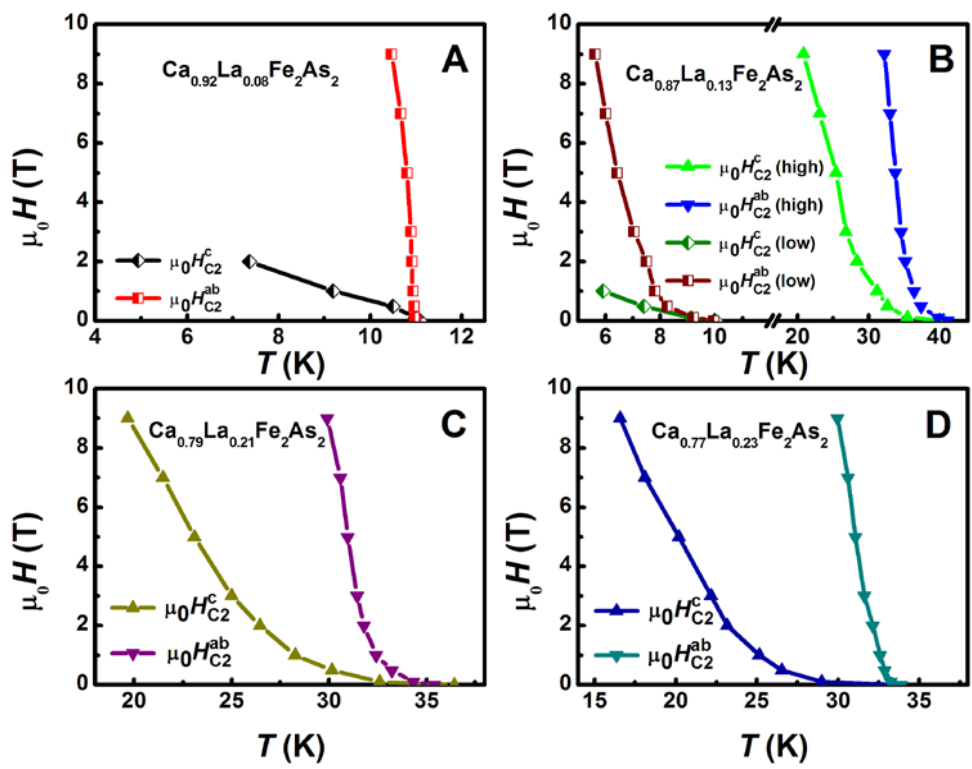

Figure 6 W. Zhou *et al*.

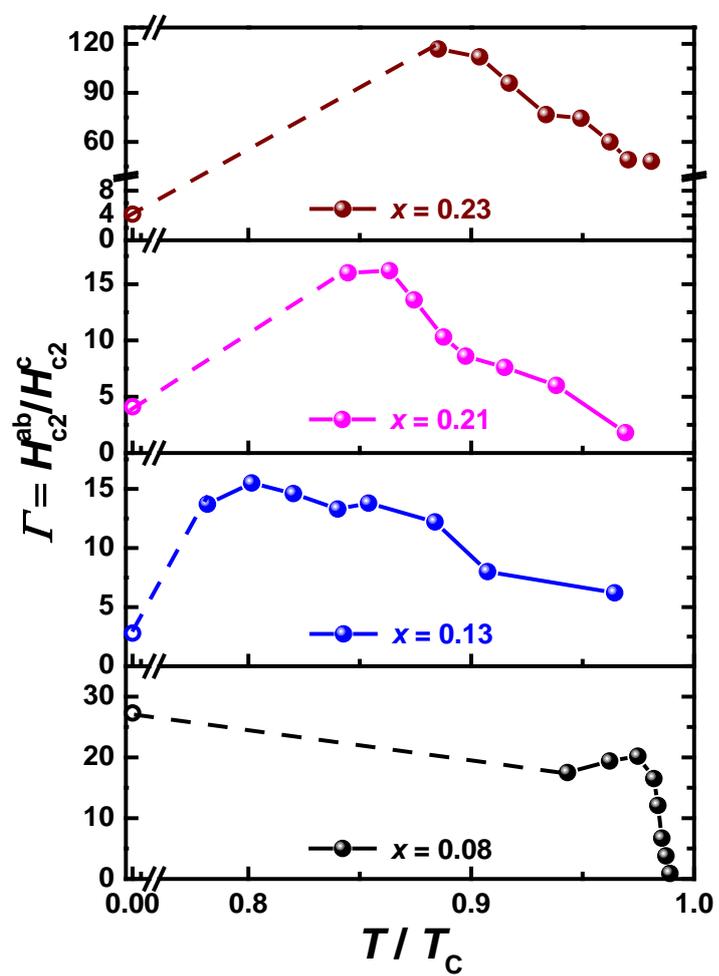

Figure 7 W. Zhou *et al*.